\documentclass{article}

\usepackage{epsfig}
\def\ligne#1{\hbox to \hsize{#1}}
\def\PlacerEn#1 #2 #3 {\rlap{\kern#1\raise#2\hbox{#3}}}
\def\cqfd{\hbox{\kern 2pt\vrule height 6pt depth 2pt width 8pt\kern 1pt}}
\def\leurre{\noindent\leftskip0pt\small\baselineskip 10pt}

\newtheorem{thm}{Theorem}

\newtheorem{fig}{Figure}
\newtheorem{tab}{Table}

\font\bfix=cmbx9

\font\ttviii=cmtt8

\font\rmix=cmr9

\def\grostrait{\ligne{\vrule height 1pt depth 1pt width \hsize}}
\def\demitrait{\ligne{\vrule height 0.5pt depth 0.5pt width \hsize}}

\begin{document}
\begin{center}
{\bf \Large An upper bound on the number of states for a strongly universal 
hyperbolic cellular automaton on the pentagrid}
\vskip 5pt
Maurice {\sc Margenstern}
\vskip 2pt
Universit\'e Paul Verlaine $-$ Metz, IUT de Metz,\\
LITA EA 3097, UFR MIM,\\
Campus du Saulcy,\\
57045 METZ C\'edex 1, FRANCE\\
{\it e-mail}: {\tt margens@univ-metz.fr}\\
{\it Web page}: {\tt http://www.lita.sciences.univ-metz.fr/\~{}margens}
\end{center}
\vskip 5pt
{\parindent 0pt\leftskip 20pt\rightskip 20pt
{\bf Abstract} $-$ In this paper, following the way opened by a previous
paper deposited on {\it arXiv}, we give an upper bound to the number of states
for a hyperbolic cellular automaton in the pentagrid. Indeed, we prove that
there is a hyperbolic cellular automaton which is rotation invariant and whose 
halting problem is undecidable and which has $9$~states.
\par}

\section{Introduction}
\label{intro}
   In~\cite{mmarXiv1DCA254}, we gave a general tool to embed a 1D-cellular automaton
into a whole family of tilings of the hyperbolic plane and into two tilings of the
hyperbolic 3D-space.

   In this paper, we try to improve the method in order to find a strongly universal
cellular automaton in the pentagrid. We remind the reader that by strong universality,
we mean a cellular automaton which mimics the computation of universal devices
starting from a {\bf finite} configuration. We also remind the reader that the pentagrid
is the tiling $\{5,4\}$, {\it i.e.} the tessellation of the hyperbolic plane based on
the regular pentagon with right angles, see~\cite{mmJUCSii,mmbook1}. 

   In Section~\ref{scenario}, we give the outline of the construction, assuming that
the reader is familiar with both cellular automata and their implementation in
tessellations of the hyperbolic plane, see~\cite{mmbook1,mmbook2}. In 
Section~\ref{propagation} we implement the preliminary structure of the implementation.
In Section~\ref{13states} we give a construction with 13~states. In 
Section~\ref{12states}, we reduce this number by one state, which will give the way
to Section~\ref{9states} where the number of states is reduced to~9 of them. 
Section~\ref{conclusion} 
concludes the paper with indications on further work.

\section{Scenario}
\label{scenario}

   In traditional literature on cellular automata, a {\bf quiescent} state
is defined as a state~$q$ such that if a cell and all its neighbours are under 
state~$q$, the cell remains under state~$q$ at the next time of the clock.
By analogy with the empty squares of the tape of a Turing machine which are said to contain
the {\bf blank} symbol, we shall fix a quiescent state and we shall call it the {\bf blank}.

For cellular automaton which have a blank state, an initial finite configuration is a
configuration in which all cells are blank except, possibly, finitely many of them.
It is plain that if the initial configuration is finite, all further configurations 
are also finite, even when the computation requires an infinite time: in this
case, there may be no bound to the size of each configuration.

   Let us now turn to the idea of the construction.

   The idea of~\cite{mmarXiv1DCA254}, which embeds any 1D-cellular automaton with
infinite initial configuration is very simple: it consists in embedding the
1D-structure of the considered cellular automaton into the desired tiling of 
the hyperbolic plane. In this construction, we simply had to devise a simple way 
to make the cells of the embedded structure different from the other cells of the 
tiling. In this way, we can easily transport the rules of the 1D-cellular automaton 
into those of the hyperbolic cellular automaton. The key point is that in this 
construction, the differentiation is made a priori: it is given in the initial 
configuration.

If we wish to implement a 1D-cellular automaton which starts its computation
from a finite configuration, then we have to go on the embedding at the same time
as the computation is going on. And so, we have to find out a simple way to
construct the 1D-structure together with the computation. But this is not enough.
Remember that the halting of the computation of a cellular automaton is defined
by the occurrence of two consecutive identical configurations. And so, when
the computation of the 1D-cellular automaton is completed, we have to stop
the construction of the 1D-structure.

   In~\cite{mmASTC}, the propagation of the tree structure of the pentagrid
is implemented in a triangular cellular automaton, and this automaton can easily 
be adapted to the pentagrid, also as a {\bf rotation invariant} one. Rotation invariance
means that the new state is unchanged if we perform a circular permutation on
the neighbours of the cell, this cell being excepted. This was done in~\cite{mmbook2} 
and repeated, as an example in~\cite{mmTCS4st}, not in the shortest way as in that
context the goal was that a cell should recognize whether it is black or white
with respect to the Fibonacci structure simply by looking at its neighbourhood.
Here, we do not really bother of this condition so that instead of six states as
it is the case in~\cite{mmbook2}, three states are enough: the blank, a white one 
and a black one. As we have the choice for the place of the initial configuration,
we can place it around the central cell which spares us the burden of initializing
the propagation of the structure: it is enough to assume it is installed in the
initial configuration and to continue it, this spares one state. 

   However, we have to stop the computation, which means that a signal has to be sent
to stop the others. In order to perform this task, it is needed to slow down the 
propagation. Indeed, the speed of a signal is at most~1. And so, if the halting signal 
travels at speed~1, it can catch up previously sent signals only if these latter
signals travelled at a lower pace. Now, slowing down necessarily costs states as we 
shall see. But, fortunately, the 1D-cellular automaton which we shall consider is
also slow, so that we shall not have to slow down too much.
 
   After that, we have to look at the way to find a 1D-cellular automaton which is
strongly universal with a small number of states. We shall use the implementation
of the $7\times4$ universal Turing Machine of Marvey Minsky which is precisely
described in~\cite{lindgren-nordahl}. This gives us a 1D-, strongly universal
cellular automaton with 7~states. In the rest of the paper, we denote this 
cellular automaton by~$\cal L$.

   Now, to see how many states we can obtain, we have to go into finer details of the
implementation. The first step is the propagation of the 1D-structure which is a common
feature of the three cellular automata which we construct in the paper.

\section{Implementation of the 1D-structure}
\label{propagation}
    In~\cite{mmarXiv1DCA254}, in the case of the pentagrid, we implemented the line of 
a 1D-cellular automaton along a line of the pentagrid. We remind the reader that such 
a line is any line which contains a side of a pentagon of the tiling: such a line 
contains the sides of infinitely many pentagons which can be gathered into two sequences
of pentagons indexed by~$Z\!\!\!Z$, two consecutive pentagons having a side on the line,
these just indicated sides being also consecutive.

    Here, as we start from a finite configuration, we have at most finitely many cells
along such a line and our task is to devise a way to go on this line as a continuation
of the segment which already exists.

\vskip 10pt
\vtop{
\vspace{-10pt}
\setbox110=\hbox{\epsfig{file=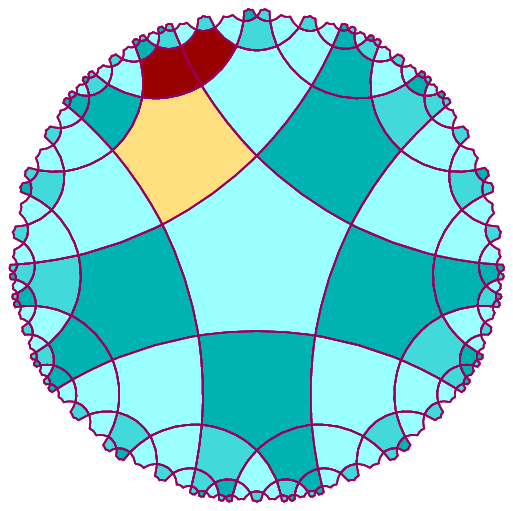,width=160pt}}
\setbox112=\hbox{\epsfig{file=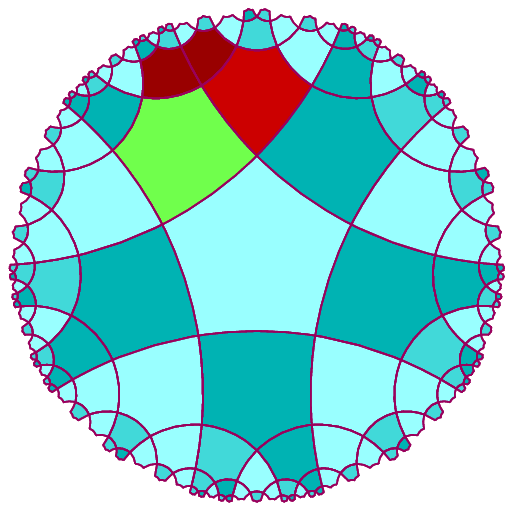,width=160pt}}
\ligne{\hfill
\PlacerEn {-162pt} {0pt} \box110
\PlacerEn {10pt} {0pt} \box112
\hfill
}
\vspace{-5pt}
\begin{fig}
\label{init_prop_0}
\leurre
The first two configurations of the propagation of the $1D$-structure.
Left-hand side: initial configuration, say time~$0$. In dark red, cells in 
state~{\bfix B},
in light yellow, the cell in state~{\bfix W}$_0$. Right-hand side: time~$1$. In right 
red, the cell in state~{\bfix B}$_0$. In green, the cell in state~{\bfix W}$_1$.
The blue cells represent the blank denoted by {\bfix N}: the different hues of blue 
remind the tree structure of the tiling, but they represent a single state.
\end{fig}
}
\vskip 10pt
    Now, the problem can be a bit simplified by the fact that $\cal L$~possesses an 
interesting feature. The cellular automaton~$\cal L$ implements a Turing machine which
is an interpreter of tag systems. This mean that we may assume that the Turing machine
works on a semi-infinite tape: {\it i.e.} the tape has an end but it is infinite in one
direction only. This is particularly interesting for our implementation: this allows us
to implement a ray only, so that we can put the end of this ray around the central cell.
For the propagation algorithm, we can define the first two configurations as 
illustrated by Figure~\ref{init_prop_0}.

\def\NN{\hbox{\bf N}}
\def\BB{\hbox{\bf B}}
\def\BZ{\hbox{\bf B$_0$}}
\def\WW{\hbox{\bf W}}
\def\WZ{\hbox{\bf W$_0$}}
\def\WU{\hbox{\bf W$_1$}}
    We need six states for the propagation of the ray: \NN, \BZ, \BB,
\WZ, \WU{} and~\WW. Informally, cells in~\BB{} will follow a branch 
of black nodes of one of the Fibonacci trees rooted around the central cell. This branch
is the leftmost branch of the chosen Fibonacci tree. Now, the cells~\WW{} follow
the rightmost branch of the next Fibonacci tree while clock-wise turning around the central
cell. It is easy to remark that the pentagons of these two branch share a ray which
supports one side of each one of these pentagons. 

   The propagation itself advances at a speed~$\displaystyle{1\over2}$. This is suggested
by the presence of the states \BZ{} and~\BB{} as well as by that of the 
states~\WZ{} and~\WU{}. 

    The mechanism is the following. A new \BB{} is produced by the 
transformation of~\BZ{} into~\BB. Now, a new~\BZ{} is obtained by the continuation of
both the black branch and the white one. It is created by the simultaneous occurrence
of a~\BB{} and a~\WZ{} around a blank cell abutting the cell at contiguous sides and in
a precise order: while counter-clockwise turning around the cell, we first meet~\BB{}
and then, immediately, \WZ. The three cells, the blank, \BB{} and \WZ{} share a common
vertex. We know that there is a fourth cell. In the initial configuration it
is a cell in~\BB. In the other configurations, when this local configuration occurs, 
the fourth cell is a cell in~\WW{}: it is a cell in~\WU{} which evolved into~\WW.
\vskip 10pt
\vtop{
\vspace{-10pt}
\setbox110=\hbox{\epsfig{file=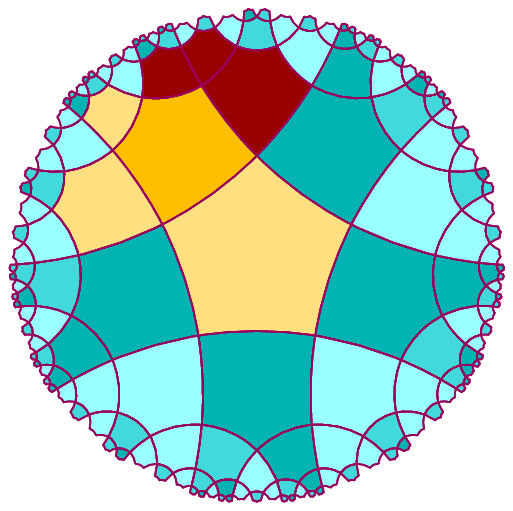,width=160pt}}
\setbox112=\hbox{\epsfig{file=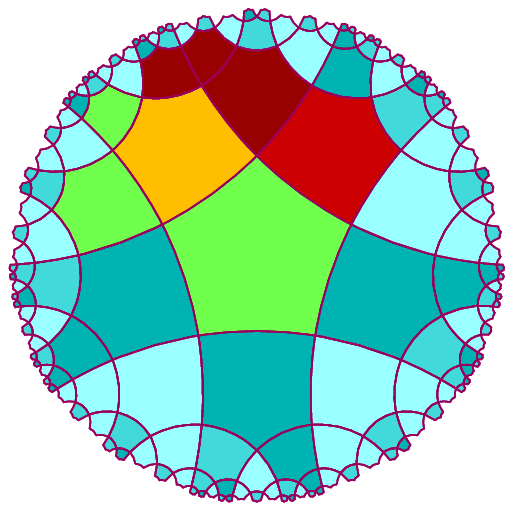,width=160pt}}
\ligne{\hfill
\PlacerEn {-162pt} {0pt} \box110
\PlacerEn {10pt} {0pt} \box112
\hfill
}
\vspace{-5pt}
\begin{fig}
\label{init_prop_1}
\leurre
Left-hand side: time~$2$ of the propagation.
Right-hand side: time~$3$ of the propagation.
In bright yellow, state~{\bf W} which indicates a fixed cell. The blue cells have the
same meaning as in Figure~{\rmix\ref{init_prop_0}}.
\end{fig}
}
\vskip 10pt
   The roles between \WZ{} and \WU{} are the following. A cell in~\WZ{} becomes~\NN{}
at the next time if it is surrounded by cells in~\NN. Otherwise, it becomes~\WU.
Now, a cell in~\WU{} becomes~\WW{} if and only if it sees a neighbour in~\BZ. Otherwise,
it becomes~\NN. Now, a cell in~\NN{} becomes~\WZ{} if and only if it has one neighbour
in~\WU{} exactly. Otherwise, it remains~\NN.

    Now, we can see that the configuration which allows a cell in~\NN{} to become~\BZ{}
requires the cells in~\WW{} or~\WZ{} to be one step in advance with respect to those 
in~\BB{} and~\BZ. This is also allowed by the progression of the cells in~\WZ. The condition
on~\WZ{} allows us to stop the progression of the~\WZ's which do not follow the ray.
Whence the importance on the condition of the transformation of the~\WZ's and also of 
the~\WU's which disappear unless they can see a cell in~\BZ, in which case they become~\WW.

\vtop{
\begin{tab}\label{t_propa}
\leurre
Rules for the propagation o the $1D$-structure.\vskip 0pt
\noindent
Left-hand side: conservative rules, marked with~$0$. Right-hand side: rules
where the current state is changed, marked with~$1$ or~$2$. The rules are rotationally
independent.
\end{tab}
\vspace{-12pt}
\grostrait
\setbox110=\vtop{\leftskip 0pt\parindent 0pt\hsize=140pt
{\ttviii
\obeylines
\leftskip 0pt
\obeyspaces\global\let =\ \parskip=-2pt
--
-- the conservative rules:
--
--     0   1   2   3   4   5   6
--
0      N   N   N   N   N   N   N
--
0      W   B   W   N   N   W   W
0      W   B   B   N   N   W   W
0      W   B   W   N   W0  W0  W
0      W   B   W   N   W1  W1  W
0      W   B   B   W0  W0  W0  W
0      W   B   B   W1  W1  W1  W
0      N   W   N   N   N   N   N
0      N   W0  N   N   N   N   N
0      N   B   N   N   N   N   N
0      N   B0  N   N   N   N   N
0      N   W   N   W0  W0  N   N
0      N   W1  W1  N   N   N   N
0      N   W0  W0  N   N   N   N
0      N   W0  B   N   N   N   N
0      N   W0  W   N   N   N   N
0      N   N   W   W0  W0  W0  N
0      N   B   N   N   N   W1  N
0      B   W0  B   N   N   N   B
0      B   B   N   N   N   N   B
0      B   W1  B   N   N   N   B
0      B   B0  N   N   N   B   B
0      B   N   N   N   B   W   B
0      B   N   N   N   W   B   B
0      B   N   N   N   B   B   B
0      B   B0  N   N   B   W   B
0      B   B   N   N   B   W   B
--
\par}
}
\setbox112=\vtop{\leftskip 0pt\parindent 0pt\hsize=145pt
{\ttviii
\obeylines
\leftskip 0pt
\obeyspaces\global\let =\ \parskip=-2pt
--
--  the propagation rules:
--
--     0   1   2   3   4   5   6
--
1      W0  N   N   N   N   N   N
1      W0  W   N   N   N   N   W1
1      W0  N   N   B   N   N   W1
1      N   B   W0  N   N   N   B0
2      B0  B   W1  N   N   N   B
1      N   W1  N   N   N   N   W0
1      W1  B0  B   N   N   N   W
1      W1  B0  W   N   N   N   W
1      W1  W   N   N   N   N   N
--
--     the format is as follows:
--
--     e0  e1  e2  e3  e4  e5  e6
--
--     e0 : current state 
--             of the cell
--     e1 : state of neighbour i,
--             seen through side i
--     e6 : new state of the cell
--             after applying the 
--             rule
--
\par}
}
\ligne{\hfill\box110 \hfill\box112 \hfill}
\vskip 7pt
\demitrait
\vskip 7pt
}
\vskip 5pt
    Table~\ref{t_propa} indicates the rules for the propagation. We can see two kinds of 
rules in the table. In the first group of rules, the current state of the cell is left
unchanged: this is why these rules are called {\bf conservative}. In the second group, the
current state of the cell is changed: these rules are called {\bf propagation} rules. 

   Basically, there are two propagation rules: the rule 
\hbox{\tt N   B   W0  N   N   N   B0}
and the rule \hbox{\tt N   W1  N   N   N   N   W0}. In both of them, the blank is changed
into a cell which will contribute to the extension of the 1D-structure. Now, the other rules
contribute to create the context required by these two propagation rules as well
as the transformation of a first signal, \BZ{} and \WZ{} into the final one, \BB{}
and~\WW, respectively.  

   The fact that we have three signals for the white node instead of two as the
speed of progression of the $1D$-structure is~$\displaystyle{1\over2}$ is explained
by the structure of the pentagrid: when a cell in~\WU{} propagates the signal~\WZ,
this is performed upon several cells, at least two of them, while it is needed for
only one of them. This is the reason of the signal~\WU{} which is an intermediate
step between \WZ{} and~\WW. Now, in order to keep the speed~$\displaystyle{1\over2}$,
the alternation is performed between~\WU{} and~\WZ.

\vskip 10pt
\vtop{
\vspace{-60pt}
\setbox110=\hbox{\epsfig{file=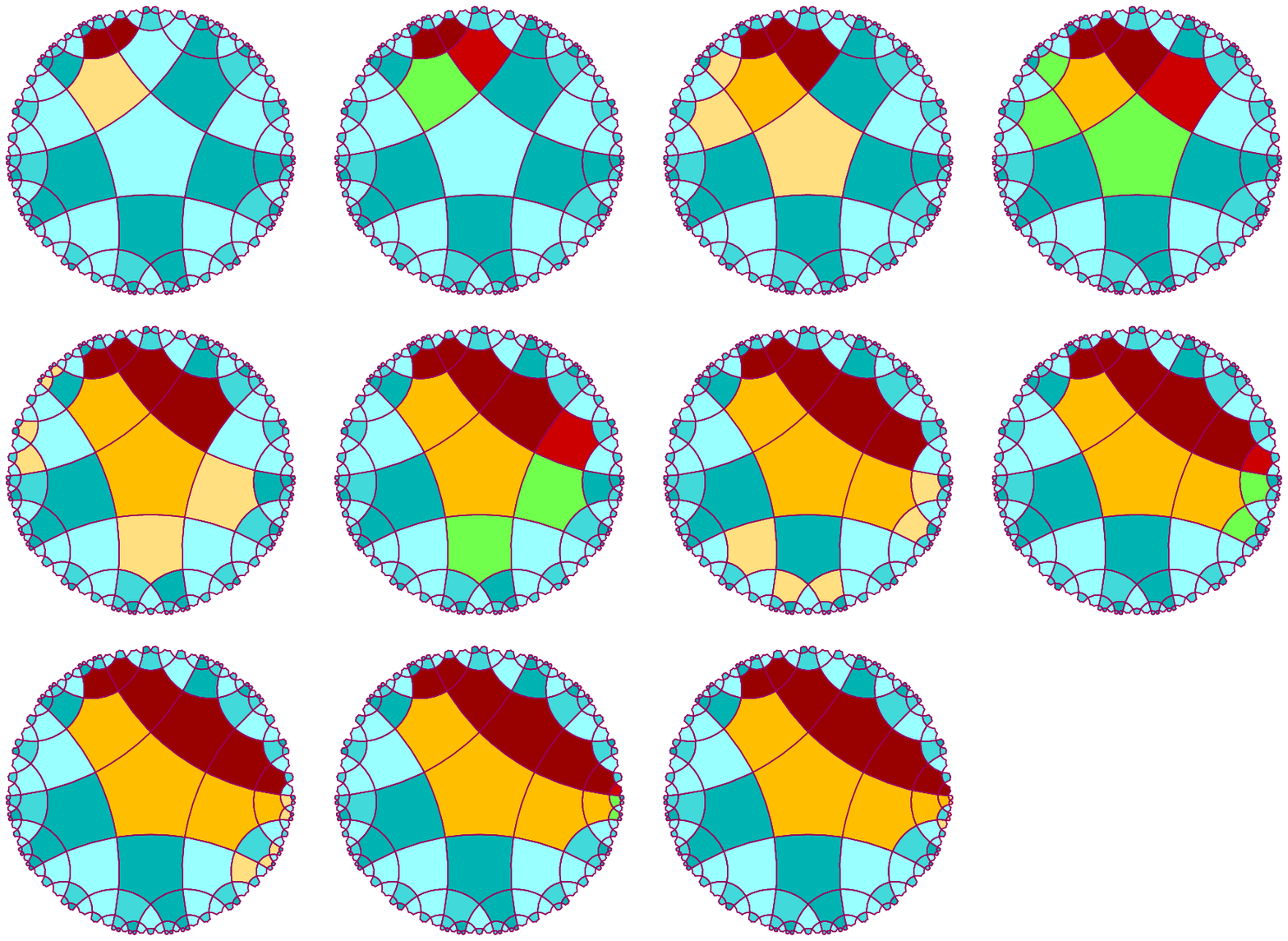,width=300pt}}
\ligne{\hfill
\PlacerEn {-126pt} {0pt} \box110
\hfill
}
\vspace{-55pt}
\begin{fig}
\label{f_propa}
\leurre
Illustration of the propagation from the beginning until a few steps.
In bright yellow, state~{\bf W} which indicates a fixed cell. The blue cells have the
same meaning as in Figure~{\rmix\ref{init_prop_0}}.
\end{fig}
}
\vskip 10pt
   It is now time to go to the other parts of the implementation. In these parts,
we shall refer to the cells in~\BB{} and in~\WW{} of the just described construction
as the {\bf ray} in which the cells in~\BB{} constitute the {\bf track} and the
cells in~\WW{} constitute the {\bf support} of the track.

\section{The 13-state cellular automaton}
\label{13states}
   First of all, we remind the reader that in this paper, we shall consider deterministic
cellular automata only, as $\cal L$~is itself a deterministic $1D$-cellular automaton.
Let us denote by~$\cal A$ the automaton which implements the computation performed
by~$\cal L$. We require $\cal A$ to be rotation invariant.

   In the previous propagation, we consider that \BB{} represents the blank of~$\cal L$
which has to be distinct from the blank~\NN{} of~$\cal A$. 
Indeed, if we give the same state for the two blanks, we shall have problems with the
propagation, as can easily be seen from the scenario of Section~\ref{scenario}.

   An important point is that in the working of~$\cal L$, it can be noticed 
from~\cite{lindgren-nordahl} that the configuration of~$\cal L$, {\it i.e.} the smallest
interval which contains the non blank cells, remains the same during at least three
consecutive steps and that it may go outside by one cell, only at the fourth time.
This means that the progression of the configuration of~$\cal L$ is much slower than the
propagation of the ray described in Section~\ref{propagation}. As a consequence, when the
signal of~$\cal L$ goes outside the current configuration, the track is ready to deliver
free blank cells.

   Accordingly, the computation of~$\cal A$ is able to perform that of~$\cal L$ during
the propagation stage. In fact, it is enough to consider that in the process described
in section~\ref{propagation}, \BB{} can be any of the states of~$\cal L$ and that
it must be the blank for the cell in~\BZ{} which is transformed into~\BB. This latter
\BB, at this moment, is the blank of~$\cal L$. Indeed, the cells of the track are the
single one which, except the ones which are at the ends of the track, have two neighbours
which are also cells of the track. In fact, except for three or four exceptional cells,
a rule of the track is of the form \hbox{\tt $y$ $x$ W $z$ N N $u$} where 
\hbox{$xyz\rightarrow u$} is a rule of~$\cal L$. For the exceptional cells, they
involve \BB{} only as a state of~$\cal L$ and, from the pictures of 
Figure~\ref{f_propa}, there cannot be ambiguity about these local configurations.

   This means that, presently, as \BB{} is one of the states of~$\cal L$, $\cal A$
has 12 states as $\cal L$ itself has 7~states, see~\cite{lindgren-nordahl}.
\vskip 5pt
   Now, let us closer look at the working of~$\cal L$. In~\cite{lindgren-nordahl},
$\cal L$~mimics rather closely the computation of Minsky's Turing machine. In particular,
there is a state~$T$, notation of~\cite{lindgren-nordahl} which represents the position
of the head of the machine. In the simulation devised in~\cite{lindgren-nordahl}, the 
halting is performed by the disappearance of~$T$. Our task is to change this
transformation into a signal which will trigger the stage at the end of which the
computation of~$\cal A$ will also halt in the traditional sense of the halting of 
a cellular automaton starting from a finite configuration, see Section~\ref{scenario}.
  
\def\blank{\hbox{\vrule height 3pt depth 0.3pt width 0.6pt
\vrule height 0.3pt depth 0.3pt width 8pt
\vrule height 3pt depth 0.3pt width 0.6pt}}

\def\HH{\hbox{\bf H}}
   To perform this task, we replace the instruction $0Ty\rightarrow 0$ of
the table of~$\cal L$ in~\cite{lindgren-nordahl} by the instruction $OTy\rightarrow \HH$, 
where \HH{} is a new state of~$\cal A$. Also, we append the new instructions
\hbox{$\HH y \BB \rightarrow \BB$}, \hbox{$\BB 0 T \rightarrow \BB$}, where \BB{} is
in both cases \blank, the blank of~$\cal L$. Note that 
\hbox{$\blank{}\, 0\, T \rightarrow \blank$} is a rule of~$\cal L$,
see~\cite{lindgren-nordahl}.

   Now, we can make a bit more precise the scenario depicted in Section~\ref{scenario}.
As just indicated, \HH{} appears on the track. In some sense it is far from the ends as 
we can put several blanks to the left of the leftmost non blank cell of the Turing tape
in the initial configuration. We remind the reader that we may choose the initial
configuration and we may choose it so that the initial segment of~$\cal L$
outside which there are only blank cells is in middle of the initial configuration, with
at least two blank cells outside this segment. We decide that \HH{} does not
affect the cells of the track which will remain unchanged. However, we decide that a
cell in~\WW{} which sees at least one~\HH{} among its neighbours becomes~\HH{} itself.

   In this way, the cells of the support of the track are progressively changed to~\HH. 
The corresponding rules are given in Table~\ref{t_propah} and they are illustrated
by Figure~\ref{f_propah}. We have just to see that we can effectively stop the 
generation of cells in~\WZ{} and in~\WU, which will to its turn stop the production 
of~\BZ. We also have to check that no problem arises on the fixed end of the ray which
represents the leftmost part of the Turing tape. 

\vtop{
\begin{tab}\label{t_propah}
\leurre
Rules for the halting of the computation.
\vskip 0pt
\noindent
Left-hand side: propagation of~{\bfix H} in the support of the track.
Right-hand side: the rules for ending the propagation of the ray and for stopping the
propagation of {\bfix H} at the origin. The rules are rotationally independent.
\end{tab}
\vspace{-12pt}
\grostrait
\setbox110=\vtop{\leftskip 0pt\parindent 0pt\hsize=140pt
{\ttviii
\obeylines
\leftskip 0pt
\obeyspaces\global\let =\ \parskip=-2pt
--
--  halting the computation:
--  spreading H
--
--     0   1   2   3   4   5   6
--
0      B   H   N   N   B   W   B
0      H   W   B   N   N   B   H
0      B   W   B   N   N   H   B
0      N   H   N   N   N   N   N
0      H   H   W   N   N   W   H
0      H   H   B   N   N   B   H
0      H   H   N   N   H   H   H
0      H   B   W   N   N   H   H
0      H   B   H   N   N   W   H
0      H   B   H   N   N   H   H
0      B   N   N   B   H   H   B
0      B   N   H   H   B   N   B
0      B   N   N   B   H   B   B
1      W   H   W   N   N   W   H
1      W   B   W   N   N   H   H
1      W   B   H   N   N   W   H
\par}
}
\setbox112=\vtop{\leftskip 0pt\parindent 0pt\hsize=140pt
{\ttviii
\obeylines
\leftskip 0pt
\obeyspaces\global\let =\ \parskip=-2pt
--
--  halting the computation:
--  end of the propagation
--
--     0   1   2   3   4   5   6
--
0      N   H   N   N   N   W1  N
0      N   N   W0  W0  N   H   N
0      N   H   N   N   N   W0  N
0      N   N   H   W0  W0  W0  N
0      N   B0  H   N   N   N   N
0      H   B   H   N   W1  W1  H
0      H   B   H   N   W0  W0  H
0      H   B   H   N   N   N   H
0      B   N   B   H   B0  N   B
0      B   N   B   H   N   N   B
1      W   B   H   W1  W1  W1  H
1      W   B   H   N   W0  W0  H
1      W1  H   N   N   N   N   N
1      W1  B0  H   N   N   N   H
1      W0  N   H   N   N   N   N
1      B0  N   B   N   N   N   N
--
--  halting the computation:
--  ending at the origin
--
0      H   B   N   N   H   B   H
0      B   B   N   N   N   H   B
1      W   N   N   H   B   B   H
\par}
}
\ligne{\hfill\box110 \hfill\box112 \hfill}
\vskip 7pt
\demitrait
\vskip 7pt
}
\vskip 5pt

   Figure~\ref{f_propah} shows how the propagation of the ray is stopped by the 
arrival of the states~\HH. For simplicity, call signal~\HH, the propagation of~\HH{}
replacing the state~\WW{} in the cells of the support of the track.

First, assume that the signal~\HH{} arrives as indicated in the figure: almost all cells of
this part of the support are now in state~\HH{}, and just a single cell in~\WW{} remains
whose next neighbour is a cell in~\WU. Necessarily, the cell in~\WW{} is changed to~\HH{}
and the cell in~\WU{} becomes~\WW: the cell in~\WU{} cannot see what is on another
side of its neighbour in~\WW. And so, at the next step, we have a cell in~\WW{} again
for which one neighbour is in~\HH{} and two others are in~\WZ. At the next time,
the cells in~\WZ{} become~\WU{} and the cell in~\WW{} becomes~\HH. Now, the cell in~\HH{}
is neighbouring a cell in~\WU. If the cell in~\WU{} turns to~\WW{} again, then the signal
will never stop the propagation of the ray. And so, the cell in~\WU{} turns to~\HH{}
when one of its neighbours is in~\HH: this is possible as no rule with the current 
state~\WU{} involved a neighbour in~\HH. And this solves the problem: at the next time,
the cells in~\WZ{} either have their five neighbours in~\NN, or they have a neighbour 
in~\HH. We decide that the neighbouring of~\HH{} makes a cell in~\WZ{} to turn to~\NN{}
too, and this stops the process. It can be checked that the rules of Table~\ref{t_propah}
allow to perform this task. This was checked by the computer program which also computed
the data for the PostScript file producing Figures~\ref{f_propa}, \ref{f_propah},
\ref{f_propaz} and~\ref{f_originh}. The computer program also checked the rotation
invariance of the rules.

\vskip 10pt
\vtop{
\vspace{-10pt}
\setbox110=\hbox{\epsfig{file=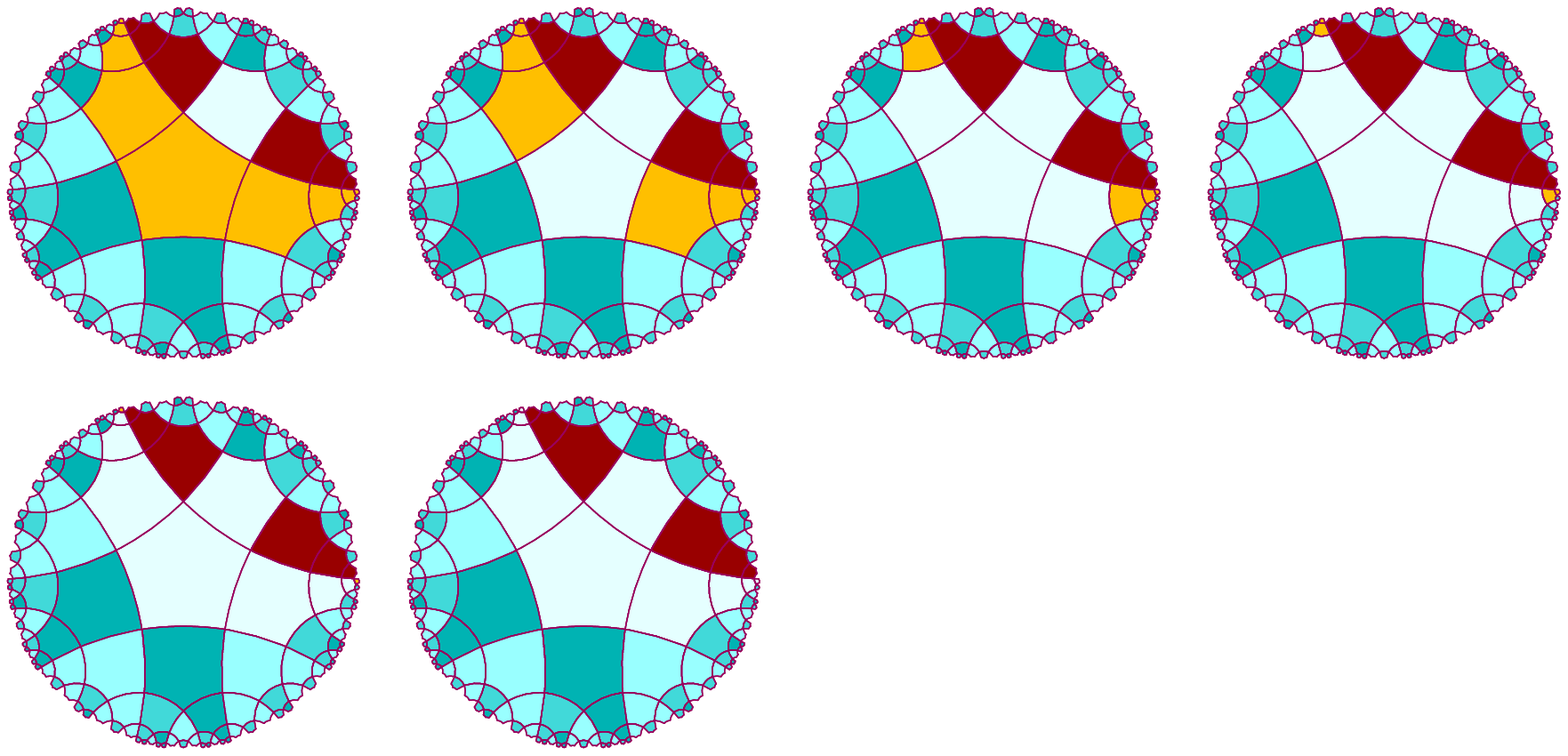,width=300pt}}
\ligne{\hfill
\PlacerEn {-126pt} {0pt} \box110
\hfill
}
\vspace{-5pt}
\begin{fig}
\label{f_propah}
\leurre
The propagation of the {\bfix H}-signal.
\end{fig}
}
\vskip 5pt
   From the figure, it is not difficult to see that the situation illustrated by 
Figure~\ref{f_propaz} is general: as the signal goes faster than the progression
of the ray, there will always be a time when the rightmost cell in~\HH{} will be close
to the rightmost cell in~\WW{} at a time when this cell is neighboured by cells in~\WU.
Indeed, if there are two cells in~\WW{} between the cell in\HH{} and the
cell in~\WU, at the next time, the cell in\WU{} which is close to the track becomes~\WW{}
while the others become~\NN{}, and the blank cells close to~\WU{} become~\WZ. Now, between
the rightmost cell in~\HH{} and the cell in~\WZ, there are two cells in~\WW. But
at the next time, the cell in~\WW{} close to~\HH{} becomes~\HH{} and the cell in~\WZ{}
which is close to the border becomes~\WU. And so, at this time, we have the same
configuration as the one illustrated by Figure~\ref{f_propaz}.
This proves that, in all cases, the signal~\HH{}
stops the progression of the ray as this was planned.

   We remain with checking that the progression of the signal~\HH{} in the other direction
is stopped by the origin: this is illustrated by Figure~\ref{f_originh}. The rules are
already given in Table~\ref{t_propah}. 

\vskip 10pt
\vtop{
\vspace{-10pt}
\setbox110=\hbox{\epsfig{file=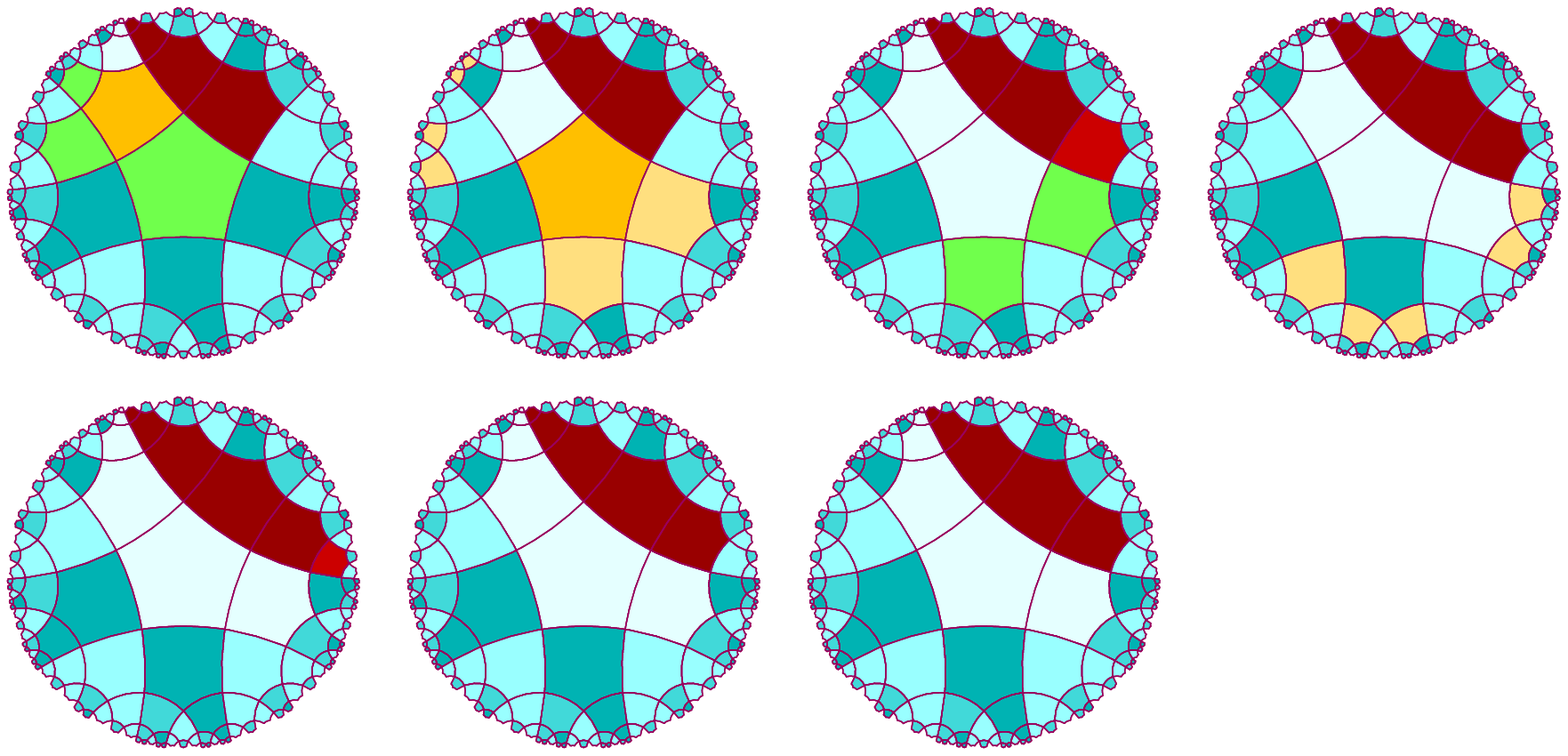,width=300pt}}
\ligne{\hfill
\PlacerEn {-126pt} {0pt} \box110
\hfill
}
\vspace{-5pt}
\begin{fig}
\label{f_propaz}
\leurre
How the {\bfix H}-signal stops the propagation of the ray.
\end{fig}
}
\vskip 5pt
\vskip 10pt
\vtop{
\vspace{-10pt}
\setbox110=\hbox{\epsfig{file=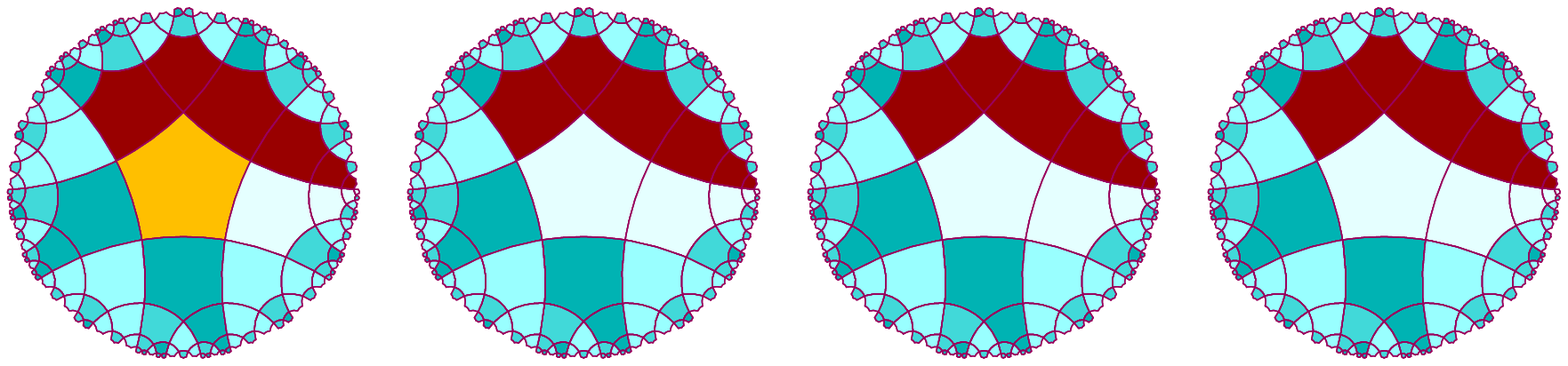,width=300pt}}
\ligne{\hfill
\PlacerEn {-146pt} {0pt} \box110
\hfill
}
\vspace{-5pt}
\begin{fig}
\label{f_originh}
\leurre
How the {\bfix H}-signal is stopped at the other end of the ray, near its origin.
\end{fig}
}

At this point, we have proved the following result:
\vskip 10pt
\begin{thm}\label{thm13states}
There is a rotation invariant hyperbolic cellular automaton in the pentagrid which 
starts from finite configurations and whose halting problem is undecidable which 
has $13$~states, the blank included.
\end{thm}
\vskip 5pt
   Indeed, $\cal A$ satisfy the statement of Theorem~\ref{thm13states}. However, the reader 
may wonder why we stated that the halting problem of~$\cal A$ is undecidable and why we 
did not state that $\cal A$ is strongly universal? This is due to a strange property
of Minsky's Turing machine with 7~states and 4~letters, a property which is inherited 
by~$\cal L$ as it closely simulates this Turing machine. The problem lies in the way
the Turing machine detects the halting of the simulated tag system. In fact, in this 
machine, when the halting production is found, the Turing machine erases its tape
so that when it stops, the content of the tape can no more be red. This 'defect' of
Minsky's machine was noticed and corrected by Rogozhin in~\cite{rogoCSJM}.
Of course, 'morally'
the machine is universal but rigorously, we can say no more than the statement of the
theorem.
\vskip 10pt
\section{The 12-state cellular automaton}
\label{12states}

    Now, we can show that a slight tuning of~$\cal A$ allows us to obtain a cellular
automaton which simulates the computation of~$\cal L$ using 12~states only.

    The idea is to replace the state~\HH{} by one of the states of~$\cal A$ which
is not used by $\cal L$.
In fact we have no choice. State~\WW{} cannot be chosen as the support of the
track consists already of cells in~\WW. Using \WW{} would require to circumvent
the support which would lead to more states. Similarly, neither \WZ{} nor~\WU{}
can be used as they contribute to continue the propagation. For the same reason,
\BZ{} is rules out and, of course, \BB{} cannot be used: this does not give a clear
signal that the computation halted. And so, the only possibility is~\NN.

   Let $\cal B$ be the cellular automaton in the pentagrid obtained from~$\cal A$
by replacing \HH{} by \NN{} in the rules of~$\cal A$. This mainly
concerns Table~\ref{t_propah}.

   We can easily see that under this replacement of \HH{} by \NN{} in the rules of 
Table~\ref{t_propah}, the rule obtained from \hbox{\tt N  H  N  N  N  W1 N} is 
in conflict with the rule we get from \hbox{\tt N  W1  N  N  N  N  W0}, as we require 
our cellular automaton to be rotation invariant. Of course, the solution is to cancel 
the rule obtained from \hbox{\tt N  H  N  N  N  W1 N}. And it works: there are no
more conflicts, we just have a few repetitions with the rules of~$\cal A$ where \HH{} 
does not occur.

   For completeness, Table~\ref{t_propahN} gives the rules for halting the computation
as the rules of Table~\ref{t_propa} are also rules for~$\cal B$.

   Also for completeness, Figures~\ref{f_propaN}, \ref{f_stoprayN} and~\ref{f_originN}
illustrate the application of the rules in conditions which may remind us 
Figures~\ref{f_propah}, \ref{f_propaz} and~\ref{f_originh}.

\vskip 10pt
\vtop{
\vspace{-10pt}
\setbox110=\hbox{\epsfig{file=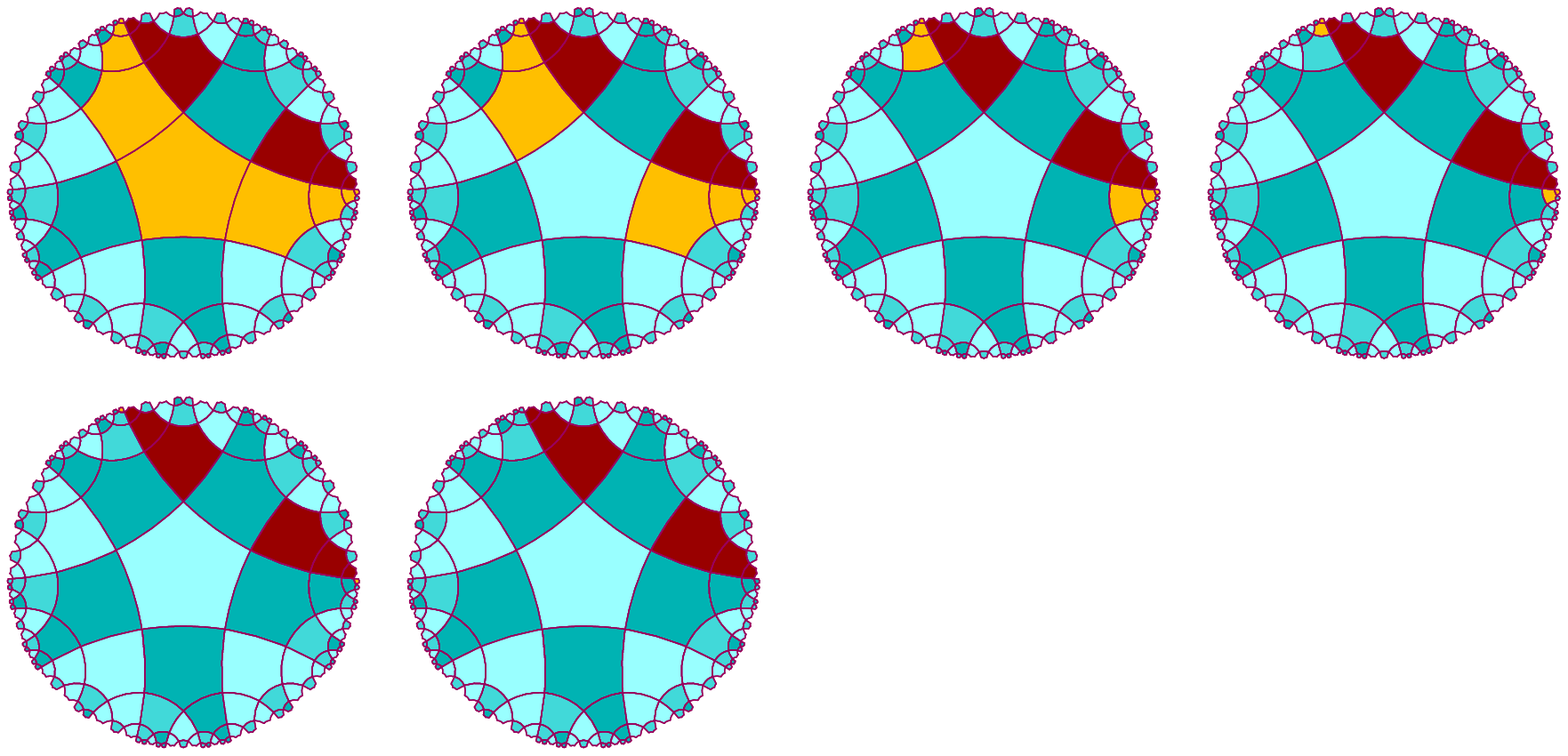,width=300pt}}
\ligne{\hfill
\PlacerEn {-146pt} {0pt} \box110
\hfill
}
\vspace{-5pt}
\begin{fig}
\label{f_propaN}
\leurre
How the {\bfix N}-signal propagates in the support of the ray: in fact, it erases 
the support.
\end{fig}
}

   And so, we proved the following result:

\vskip 10pt
\begin{thm}\label{thm12states}
There is a rotation invariant hyperbolic cellular automaton in the pentagrid which 
starts from finite configurations and whose halting problem is undecidable which 
has $12$~states exactly, the blank included.
\end{thm}

   Note that the remark about strong universality for Theorem~\ref{thm13states}
also holds for this one.

\vskip 10pt
\vtop{
\begin{tab}\label{t_propahN}
\leurre
Rules of $\cal B$ for the halting of the computation.
\vskip 0pt
\noindent
Left-hand side: propagation of~{\bfix N} in the support of the track.
Right-hand side: the rules for ending the propagation of the ray and for stopping the
propagation of {\bfix N} at the origin. The rules are rotationally independent. Rules
which were rotationally similar to rules of Table~{\rmix\ref{t_propa}} have been 
removed.
\end{tab}
\vspace{-12pt}
\grostrait
\setbox110=\vtop{\leftskip 0pt\parindent 0pt\hsize=140pt
{\ttviii
\obeylines
\leftskip 0pt
\obeyspaces\global\let =\ \parskip=-2pt
--
--  erasing the support of
--     the track
--
--     0   1   2   3   4   5   6
--
0      N   W   B   N   N   B   N
0      N   N   W   N   N   W   N
0      N   N   B   N   N   B   N
0      N   B   W   N   N   N   N
0      N   B   N   N   N   W   N
0      B   N   N   B   N   B   B
1      W   N   W   N   N   W   N
1      W   B   W   N   N   N   N
1      W   B   N   N   N   W   N
\par}
}
\setbox112=\vtop{\leftskip 0pt\parindent 0pt\hsize=140pt
{\ttviii
\obeylines
\leftskip 0pt
\obeyspaces\global\let =\ \parskip=-2pt
--
--   the rules for stopping
--      the ray
--
0      N   N   N   W0  W0  W0  N
0      N   B   N   N   W1  W1  N
0      N   B   N   N   W0  W0  N
0      N   B   N   N   W0  N   N
0      B   N   B   N   B0  N   B
1      W   B   N   W1  W1  W1  N
1      W   B   N   N   W0  W0  N
1      W1  N   N   N   N   N   N
1      W1  B0  N   N   N   N   N
1      W0  N   N   N   N   N   N
1      B0  N   B   N   N   N   N
--
--  the rules for stopping 
--     the erasing at the origin
--
0      N   B   N   N   N   B   N
1      W   N   N   N   B   B   N
\par}
}
\ligne{\hfill\box110 \hfill\box112 \hfill}
\vskip 7pt
\demitrait
\vskip 7pt
}
\vskip 10pt
\vtop{
\vspace{-10pt}
\setbox110=\hbox{\epsfig{file=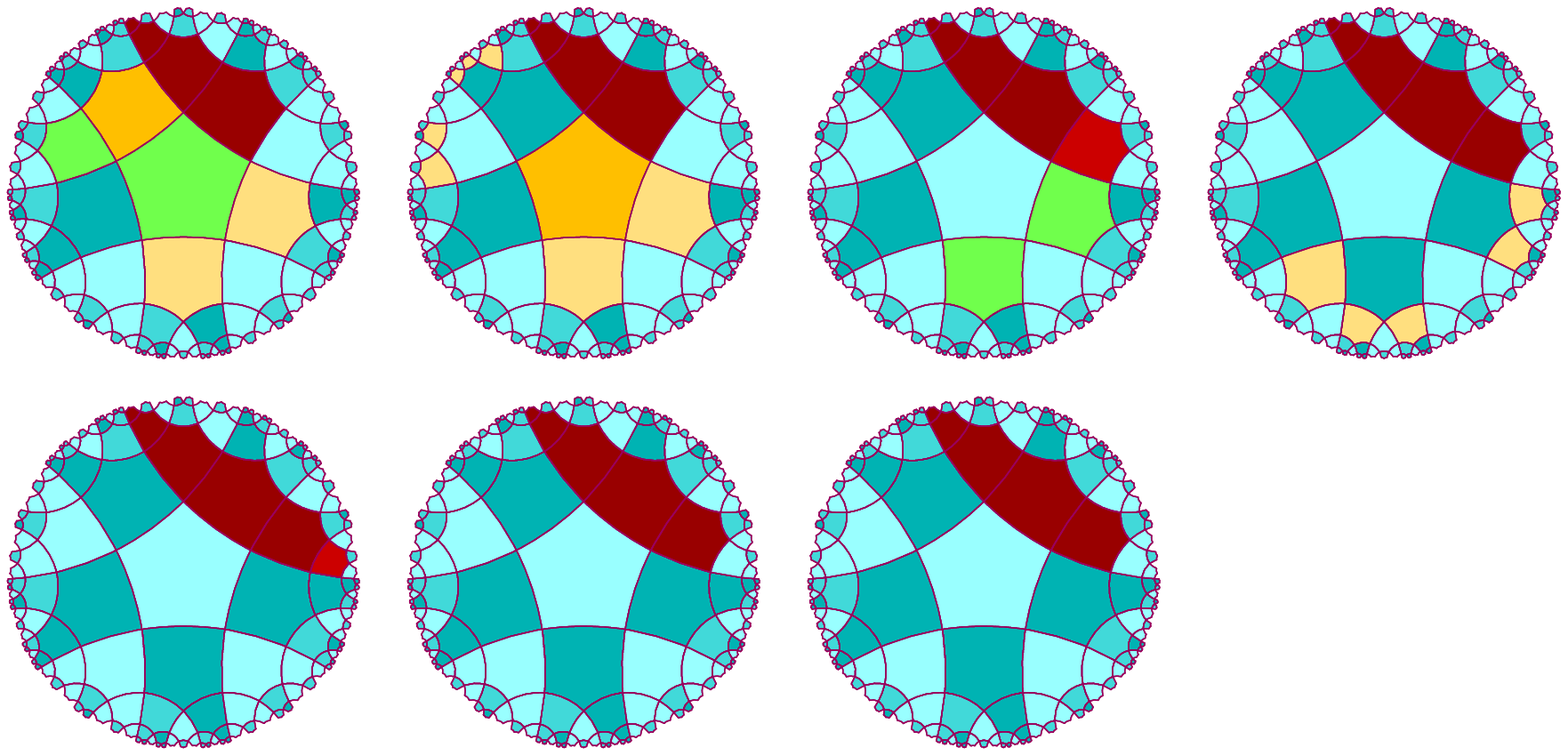,width=300pt}}
\ligne{\hfill
\PlacerEn {-146pt} {0pt} \box110
\hfill
}
\vspace{-5pt}
\begin{fig}
\label{f_stoprayN}
\leurre
How the {\bfix N}-signal stops the propagation of the ray: it completes the erasing of 
the support at this end of the ray.
\end{fig}
}

   From Figures~\ref{f_propaN}, \ref{f_stoprayN} and~\ref{f_originN}, we can notice 
that replacing \HH{} by~\NN{} boils down to erase the support of the track so that, 
at the end of the computations, we remain with the track only. This erasing occurs at
both ends of the ray as proved by Figures~\ref{f_stoprayN} and~\ref{f_originN}.
Figure~\ref{f_stoprayN} also shows that the erasing process stops the propagation of the
ray.
\vskip 10pt
\vtop{
\vspace{-20pt}
\setbox110=\hbox{\epsfig{file=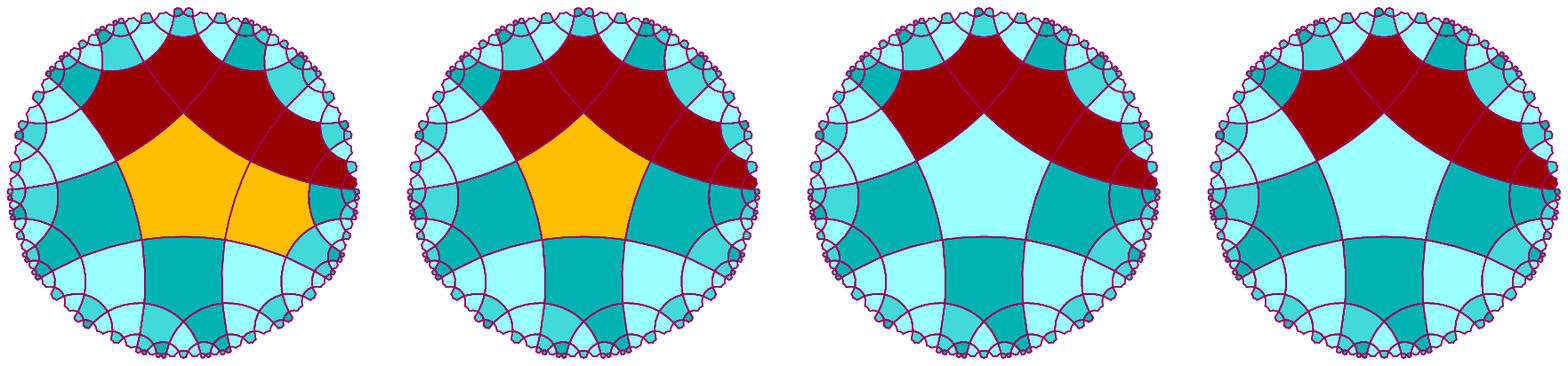,width=300pt}}
\ligne{\hfill
\PlacerEn {-146pt} {0pt} \box110
\hfill
}
\vspace{-5pt}
\begin{fig}
\label{f_originN}
\leurre
How the {\bfix N}-signal is stopped at the other end of the ray, near the origin.
\end{fig}
}

   Now, could we reduce again the number of states, using the same $1D$~cellular 
automaton? The next Section gives a positive answer to this question.

\vskip 10pt
\section{The 9-state cellular automaton} 
\label{9states}

   We have already seen that to reduce the number of states from~13 down to~12,
we replaced the state~\HH{} by the state~\NN. We can try to go further in this direction.
In~\cite{mmarXiv1DCA254}, we reduced the number of states for a weakly universal 
cellular automaton from 3~states down to 2~ones by replacing the extra state in the 
simulation with 3~states by a state of the embedded 2-states cellular automaton. So that
here, a natural idea is to replace as many as we can states from~\NN, \WW, \WZ, \WU{} 
and~\BZ, by states of~$\cal L$ only. If we look at the rules displayed 
in~\cite{lindgren-nordahl}, we can notice that the symbol~$T$ has a rather empty 
sub-table. In particular, there is no rule assigned to $TTT$, so that we can decide 
that $TTT\rightarrow T$ is used by our new automaton, $\cal C$. Inspired by the table 
of the rules given in~\cite{lindgren-nordahl}, we shall see that \WW, \WZ{} and~\BZ{} 
can be replaced by~$T$, 0 and~$A$ respectively. It is enough to check that performing these
replacements and taking the above figures, we obtain new rules which are rotation
invariant and compatible. 

   Indeed, consider the first stage of the working of~$\cal B$ which consists in
propagating the structure at the same time when the computation is going on.
It will be enough to ensure that a pure propagation process can be performed under
the new set of states and that there is no contradiction between the new involved rules and
the rules derived from the computation of~$\cal L$. We also have to check that the
new rules do not disturb the computation itself. 

   Now, this latter condition entails that we cannot replace \WU{}
by a state of~$\cal L$. Indeed, imagine that \WU{} is a state of~$\cal L$, say~$\alpha$. 
There are occurrences of~$\alpha$ on the track. Each cell of the track has at least
two neighbours in state~\NN. Consider one of these neighbours. It has $\alpha$~as a 
neighbour and all the others are in~\NN. From Section~\ref{propagation}, we know that 
in this case, the state~\NN{} is replaced by~\WZ{} and this \WZ, as it is not surrounded
by cells in~\NN{} only, will become~\WU. Now, if \WU{} has a neighbour in~$A$, which may
happen in the track, this \WU{} becomes~$T$ which will disturb the computation. Even 
in the case it will not disturb the computation, the production of~\WZ{} nearby the track
will be repeated periodically even when the computation has stopped. So that we cannot
replace \WU{} by a state of~$\cal L$. Can we replace \WZ{} by a state of~$\cal L$?
The answer is yes: in the propagation context, a cell in~\WZ{} has at least four neighbours
in~\NN. A cell of the track which is not the blank has two neighbours in~\NN{} exactly, so
that it is possible to distinguish the role of a distinguished~$\alpha$, depending on 
its neighbourhood. Moreover, a neighbour in~\NN{} of a cell in~$\alpha$ would remain
unchanged due to the rules \hbox{\NN\ \NN\ \NN\ \NN\ \NN\ \WZ\ \NN}
and \hbox{\NN\ \NN\ \NN\ \NN\ \NN\ \BB\ \NN}.

\subsection{The propagation of the ray}

   Let us have a new inspection of Figure~\ref{f_propa}.
Below, Table~\ref{releve_propa} indicates the rules applied for going from one
picture of the figure to the next one. As can be seen from the table, after time~7,
no new rule is needed for the propagation of the ray. We can notice that these rules 
are obtained from those defined in Table~\ref{t_propa} by replacing \WW, \WZ{} 
and \BZ{} by $T$, 0 and~$A$ respectively.

   However, we have to keep in mind that, during the propagation, the computation
is going on. And so, we have to look at the cells of the track and of their neighbours
in order to check that they are compatible with the rules found in 
Table~\ref{releve_propa}.

   A rule which applies to a cell of the track is of the form 
\vskip 5pt
\ligne{\hfill {\BB \NN \NN \BB\ $T$ \BB \BB}\hfill$(\ast)$.\hfill} 
\vskip 5pt
\noindent
As we know, \BB{} is a generic name for the states
of~$\cal L$. If we replace \BB{} by the states of~$\cal L$, we get rules of the form
\vskip 5pt
\ligne{\hfill $\alpha_0\NN\NN\alpha_{-1}T\alpha_1\alpha^1_0$\hfill$(\ast\ast)$\hskip 20pt} 
\vskip 5pt
\noindent
where 
$\alpha_{-1}\alpha_0\alpha_1\rightarrow\alpha^1_0$ is a rule of~$\cal L$. 
When $\alpha_0\in\{B,y\}$, then the rule 
cannot be confused with one of~$\cal C$ which we have already defined. At the times
up to~5, when \BB{} is the current state, it is always \blank, the blank of~$\cal L$. 
Now the rules of Table~\ref{releve_propa} where the current state is~\BB{} and which 
are used up to time~6 either contain at least three consecutive occurrences of~\NN, 
or they contain an occurrence of~$A$ and so, they cannot be confused with a 
rule~$(\ast\ast)$ whose current state would be \blank{}. We have to look at
all the other possibilities.

   Consider the case when \BB{} is $T$: the rules of Table~\ref{releve_propa}
where the current state is~$T$ are those of the cell 1(1) at times~3, 4 and~5, and those
of cell~0 at times~5 and~6. Only one rule contains two consecutive occurrences of~\NN:
the rule \hbox{$T$ \NN{} \NN{} $T$ \BB{} \BB{} $T$}. Note that the relative positions of 
these two occurrences of~\NN{} and that of~$T$ are fixed. Now, when $T$~occurs in a cell
of the track, it is the single occurrence of this symbol on the track as this is the case
with~$\cal L$. In particular,  in~$(\ast\ast)$,
if $\alpha_0=T$, then
$\alpha_1$ or~$\alpha_{-1}$ must be \blank{} and the other symbol is any one of~$\cal L$
except the blank and~$T$: see the table of the rules of~$\cal L$ 
in~\cite{lindgren-nordahl}. Consequently, a rule of the track when the current state 
is~$T$ cannot be confused with the rules with~$T$ as the current state in 
Table~\ref{releve_propa}.

  Now, consider the case when \BB{} is~0 or~$A$. We compare $(\ast\ast)$ with
the rules of Table~\ref{releve_propa} which have the same symbol as the current state.
When it is~$A$, the rules of Table~\ref{releve_propa} have three consecutive occurrences
of~\NN{} while there are only two of them in the rule $(\ast\ast)$. For symbol~0, we
have a similar argument: the rules~$(\ast\ast)$ with~0 as the current state have
at least three consecutive neighbours in~\NN.
\def\unerangee #1 #2 regle #3 #4 #5 #6 #7 #8 #9 {%
\ligne{\hbox to 20pt{\hfill\tt #1\hfill}
       \hbox to 30pt{\hfill#2\ :}\hskip 10pt
       \hbox to 91pt{#3\ #4\ #5\ #6\ #7\ #8\ #9\ \hfill} 
      \hfill}
} 
\def\unerangeeb #1 #2 comment #3 {%
\ligne{\hbox to 20pt{\hfill\tt #1\hfill}
       \hbox to 30pt{\hfill#2\ :}\hskip 10pt
       \hbox to 91pt{#3\hfill} 
      \hfill}
} 

\vtop{
\begin{tab}\label{releve_propa}
\leurre
Table of the rules needed for the propagation of the ray. In the table, $\nu_i$
with or without indices indicates neighbours of a cell under state~$1$ which
become under state~$0$. More precisely, $\nu_1$ consists of the cells $4(1)$,
$2(2)$ and~$0$, $\nu'_1$ consists of the cells~$4(1)$ and $2(2)$. Similarly,
$\nu_2$ consists of the cells $11(1)$, $12(1)$, $6(2)$, $7(2)$, $1(3)$ and
$1(4)$. It splits into $\nu'_2$, which consists of $11(1)$, $12(1)$, $6(2)$, $7(2)$
and $\nu"_2$, which consists of $1(3)$ and $1(4)$. At last, $\nu_3$ consists
of the cells $2(3)$, $3(3)$, $4(3)$, $3(4)$ and $4(4)$. After time~$7$, no new rule
appear in connection with the propagation of the ray.
\end{tab}
\vspace{-12pt}
\grostrait
\vskip 6pt
\ligne{\hfill
\vtop{\leftskip 0pt\parindent 0pt\hsize=165pt 
\unerangee 1 {3(1)} regle {\BB} {\NN} {\NN} {\NN} {0} {\BB} {\BB}
\unerangee {} {7(1)} regle {\BB} {\NN} {\NN} {\NN} {\NN} {\BB} {\BB}
\unerangee {} {1(1)} regle {0} {\NN} {\NN} {\NN} {\NN} {\BB} {\WU}
\unerangee {} {2(1)} regle {\NN} {\NN} {\NN} {\NN} {\BB} {0} {$A$}
\vskip 2pt
\unerangee 2 {3(1)} regle {\BB} {\NN} {\NN} {\NN} {1} {\BB} {\BB}
\unerangee {} {7(1)} regle {\BB} {\NN} {\NN} {\NN} {\BB} {$A$} {\BB}
\unerangee {} {1(1)} regle {\WU} {\NN} {\NN} {\NN} {$A$} {\BB} {$T$}
\unerangee {} {2(1)} regle {$A$} {\NN} {\NN} {\NN} {\BB} {\WU} {\BB}
\unerangee {} {$\nu_1$} regle {\NN} {\NN} {\NN} {\NN} {\NN} {\WU} {0}
\vskip 2pt
\unerangee 3 {3(1)} regle {\BB} {\NN} {\NN} {\NN} {$T$} {\BB} {\BB}
\unerangee {} {7(1)} regle {\BB} {\NN} {\NN} {\NN} {\BB} {\BB} {\BB}
\unerangee {} {1(1)} regle {$T$} {0} {0} {0} {\BB} {\BB} {$T$}
\unerangee {} {2(1)} regle {\BB} {\NN} {\NN} {\NN} {\BB} {$T$} {\BB}
\unerangee {} {$\nu_1$} regle {0} {\NN} {\NN} {\NN} {\NN} {$T$} {\WU}
\unerangeeb {} {1(5)} comment {rule for 2(1), \tt 1} 
\vskip 2pt
\unerangee 4 {1(1)} regle {$T$} {1} {1} {1} {\BB} {\BB} {$T$}
\unerangee {} {2(1)} regle {\BB} {\NN} {\NN} {\BB} {$T$} {$A$} {\BB}
\unerangee {} {$\nu'_1$} regle {1} {\NN} {\NN} {\NN} {\NN} {$T$} {\NN}
\unerangee {} {0} regle {1} {\NN} {\NN} {\NN} {$A$} {$T$} {$T$}
\unerangee {} {$\nu_2$} regle {\NN} {\NN} {\NN} {\NN} {\NN} {\WU} {0}
\unerangeeb {} {1(5)} comment {rule for 2(1), \tt 2} 
}
\hfill
\vtop{\leftskip 0pt\parindent 0pt\hsize=165pt 
\unerangee 5 {1(1)} regle {$T$} {\NN} {\NN} {$T$} {\BB} {\BB} {$T$}
\unerangee {} {2(1)} regle {\BB} {\NN} {\NN} {\BB} {$T$} {\BB} {\BB}
\unerangeeb {} {1(5)} comment {rule for 2(1), \tt 3} 
\unerangee {} {0} regle {$T$} {\NN} {0} {0} {\BB} {$T$} {$T$}
\unerangee {} {$\nu'_2$} regle {0} {\NN} {\NN} {\NN} {\NN} {\NN} {\NN}
\unerangee {} {$\nu"_2$} regle {0} {\NN} {\NN} {\NN} {\NN} {$T$} {1}
\unerangeeb 6 {1(5)} comment {rule for 2(1), \tt 4} 
\unerangeeb {} {1(3)} comment {rule for $\nu'_1$, \tt 4} 
\unerangeeb {} {1(4)} comment {rule for 0, \tt 4} 
\unerangee {} {0} regle {$T$} {\NN} {\WU} {\WU} {\BB} {$T$} {$T$}
\unerangeeb {} {$\nu_3$} comment {rule for $\nu_2$, \tt 4} 
\unerangeeb 7 {1(5)} comment {rule for 2(1), \tt 5}
\unerangee {} {0} regle {$T$} {\NN} {\NN} {$T$} {\BB} {$T$} {$T$}
\unerangee {} {1(3)} regle {\NN} {\NN} {$T$} 0 0 0 {\NN}
\unerangeeb {} {1(4)} comment {rule for 0, \tt 5} 
}
\hfill}
\vskip 6pt
\demitrait
\vskip 6pt
}  
\vskip 10pt

   We remain with the case when \BB{} is the blank of~$\cal L$. From the previous cases, 
we know that there is no possible confusion when the rule contains at least three
consecutive occurrences of~\NN. Now, two rules have two consecutive occurrences of~\NN{}
exactly: the rule \hbox{\BB\ \NN\ \NN\ \BB\ $T$\ $A$\ \BB} and the rule
\hbox{\BB\ \NN\ \NN\ \BB\ $T$\ \BB\ \BB}. These rules can be seen as rules
of the form~$(\ast\ast)$ when \hbox{\BB = \blank}. The corresponding rules
are \hbox{\blank\ \NN\ \NN\ \blank\ $T$\  $A$\ \blank} and
\hbox{\blank\ \NN\ \NN\ \blank\ $T$\  \blank\ \blank}, respectively.
Now, interpreted as rules induced by a rule of~$\cal L$, the corresponding
rules of~$\cal L$ would be \hbox{\blank$\,$\blank$\,A$\ $\rightarrow$\ \blank}
and \hbox{\blank$\,$\blank$\,$\blank\ $\rightarrow$\ \blank}. Now, these rules
are indeed present in the table of the rules of~$\cal L$, see~\cite{lindgren-nordahl}.
And so, at this stage, there is no confusion by replacing \WW, \WZ{} and \BZ{}
by $T$, 0~and~$A$ respectively.

\subsection{The erasing of the support of the track}

   We have to look at the final stage of the process. When the halting is met, \NN{}
is introduced onto the track and, as we know from Section~\ref{12states}, this
starts the erasing process of the support of the track by propagation of~\NN{}
which successively replaces all occurrences of~$T$ and, at the end of the propagation of
the ray, which stops the production of cells in~0.

   Looking at Figures~\ref{f_propaN} and~\ref{f_stoprayN}, we can see that this
requires the following rules: 

\vskip 10pt
\vtop{
\grostrait
\vskip 6pt
\ligne{\hfill
\vtop{\leftskip 0pt\parindent 0pt\hsize=165pt 
\unerangee {} 0 regle {$T$} {\NN} {\NN} {$T$} {\NN} {$T$} {\NN}
\unerangee {} 0 regle {\NN} {\NN} {\NN} {$T$} {\NN} {$T$} {\NN}
\unerangee {} {1(1)} regle {$T$} {\NN} {\NN} {\NN} {\BB} {$T$} {\NN}
\unerangee {} {1(4)} regle {$T$} {\NN} {\NN} {\NN} {$T$} {\BB} {\NN}
}
\hfill
\vtop{\leftskip 0pt\parindent 0pt\hsize=165pt 
\unerangee {} {} regle {$T$} {\NN} {\NN} {\WU} {\WU} {\BB} {\NN}
\unerangee {} {} regle {$T$} {\NN} {\NN} {0} {0} {\BB} {\NN}
\unerangee {} {} regle {\NN} {\NN} {\NN} {\NN} {\BB} {$T$} {\NN}
\unerangee {} {} regle {\NN} {\NN} {\NN} {\NN} {$T$} {\BB} {\NN}
}
\hfill}
\vskip 6pt
\demitrait
}
\vskip 10pt
   Now, it is easy to see that none of them cannot be confused with the
rule \hbox{$T$\ \NN\ \NN\ $T$\ \BB\ $T$\ $T$}  nor a rule of the form~$(\ast\ast)$ as 
these latter rules have only two occurrences of~\NN. It can also be seen that the above 
rules cannot be confused with any of the other rules of Table~\ref{releve_propa}
where the current state is~$T$: again the number of occurrences of~\NN{} is different.
Indeed, this number is two or three in the above rules while it 
is at most one only in the rules of Table~\ref{releve_propa}, except the rules
\hbox{$T$\ \NN\ \NN\ $T$\ \BB\ \BB\ $T$} and \hbox{$T$\ \NN\ \NN\ $T$\ \BB\ $T$\ $T$}.
But there can be no confusion with these latter rules either: they have~$T$
after the two occurrences of~\NN{} while in the above rules with two occurrences of~\NN{}
exactly, which are also consecutive, there is~0 or~\WU{} after the second~\NN.

   We also have to check that the track is not disturbed by the replacement of~$T$ 
by~\NN. Indeed, this replacement has, as a consequence, that the form $(\ast\ast)$
is replaced by the following one:
\vskip 5pt
\ligne{\hfill $\alpha_0\NN\NN\alpha_{-1}\NN\alpha_1\alpha^1_0$\hfill%
$(\ast\ast\ast)$\hskip 20pt} 
\vskip 5pt
\noindent
As the three occurrences of~\NN{} in $(\ast\ast\ast)$ are not consecutive, there is 
no confusion with the rules of Table~\ref{releve_propa} where the current state is~\BB.

   This completes the proof that $\cal C$ exactly simulates the computation of~$\cal L$
with a true stopping of the cellular automaton in the case when the computation
of~$\cal L$ also stops. Accordingly we have proved the following result:

\begin{thm}\label{thm9states}
There is a rotation invariant hyperbolic cellular automaton in the pentagrid which 
starts from finite configurations and whose halting problem is undecidable which 
has $9$~states exactly, the blank included.
\end{thm}

\section{Conclusion}
\label{conclusion}
   While stating Theorem~\ref{thm13states}, we have explained why we did not say that 
the cellular automaton~$\cal A$ is strongly universal and the same explanation holds
for the automata $\cal B$ and~$\cal C$ of Theorems~\ref{thm12states} 
and~\ref{thm9states} respectively.

   Can we still have a strongly universal cellular automaton with 9~states or possibly 
less?

   One way to solve this problem would be to apply the technique of~\cite{lindgren-nordahl}
to another small Turing machine. In~\cite{rogoCSJM} where the defect of this machine was
first noticed, the author provides another Turing machine with 7~states and 4~letters
which mimics any tag system of a given family, the same as for Minsky's machine, and
the machine of~\cite{rogoCSJM} is actually universal. Another Turing machine with 7~states
and 4~letters which is truly universal was later provided by R. Robinson, 
see~\cite{robinsonTM}. It would be interesting to see whether a smaller machine, as
the one devised by T.~Neary and D.~Woods with 6~states and 4~letters, 
see~\cite{neary_woods}, could yield a better solution.
 
   Accordingly, there is some work ahead, probably a tedious one if not more difficult.

\end{document}